\documentclass[conference]{IEEEtran}
\IEEEoverridecommandlockouts
\usepackage{cite}
\usepackage{amsmath,amssymb,amsfonts}
\usepackage{algorithmic}
\usepackage{graphicx}
\usepackage{textcomp}
\usepackage{xcolor}
\usepackage{makecell}
\usepackage{multirow}
\usepackage{booktabs}   
\usepackage{makecell}
\usepackage{pifont}   
\usepackage[export]{adjustbox} 
\usepackage{tabularx}      
\usepackage{xcolor}        
\usepackage{array}         
\usepackage[hidelinks]{hyperref}

\newcolumntype{L}{>{\raggedright\arraybackslash}X}

\def\BibTeX{{\rm B\kern-.05em{\sc i\kern-.025em b}\kern-.08em
    T\kern-.1667em\lower.7ex\hbox{E}\kern-.125emX}}
\begin{document}

\title{Multimodal Aspect-Level Sentiment Analysis Based on Gated Noise Filtering and Emotion-Relevance Interaction}

\author{
\IEEEauthorblockN{
Chen Huang\textsuperscript{1,2,3},
Liangwei Guo\textsuperscript{1,2,3},
Yamin Li\textsuperscript{1,2,3,*},
Yan	Zhang\textsuperscript{1,2,3},
Chao Yang\textsuperscript{1,2,3},
Li	Yang\textsuperscript{1,2,3},
Jianhua	Song\textsuperscript{4}
}
\IEEEauthorblockA{
\textsuperscript{1}School of Computer Science, Hubei University, Wuhan 430062, China \\
\textsuperscript{2}Key Laboratory of Intelligent Sensing System and Security (Hubei University), Ministry of Education, Wuhan 430062, China \\
\textsuperscript{3}Hubei Key Laboratory of Big Data Intelligent Analysis and Application (Hubei University), Wuhan 430062, China \\
\textsuperscript{4}School of Cyberspace Security,Hubei University,Wuhan 430062,China \\
}
\thanks{\textsuperscript{*}Corresponding author: Yamin Li, yamin.li@hubu.edu.cn}
}

\maketitle

\begin{abstract}
Multimodal Aspect-Based Sentiment Analysis (MABSA) infers fine-grained sentiment polarity toward specific aspects by jointly modeling text and images. Despite progress in cross-modal fusion, two challenges remain in multi-aspect settings: (1) multimodal noise, where aspect-irrelevant content distracts sentiment learning; and (2) weak cross-modal sentiment alignment, as visual evidence can be ambiguous and textual--visual sentiments may conflict, limiting multimodal complementarity. To address these issues, we propose a Gated Noise-filtered Sentiment-Relevance Interaction (GNSRI) framework. It employs a gated noise-filtering module to suppress sentiment-irrelevant features and enhance aspect-aware sentiment cues, and a sentiment-relevance interaction module to capture consistent and conflicting cross-modal signals at micro and macro levels. Finally, a learnable decision fusion mechanism adaptively combines predictions from textual, visual, and cross-modal branches at the aspect level. Experiments on public MABSA benchmarks show that GNSRI outperforms state-of-the-art methods, improving accuracy by 1.94\% and 2.06\% on Twitter-2015 and Twitter-2017, respectively.
\end{abstract}


\section{Introduction}
With the rapid growth of social media and online review platforms, user-generated content has expanded dramatically, posing new challenges to sentiment analysis. As expressions become increasingly diverse and semantically complex, traditional text-only methods are insufficient for modeling fine-grained aspect-level sentiment associations. In this context, \emph{Multimodal Aspect-Based Sentiment Analysis} (MABSA)~\cite{ref1} has emerged as an active research area, aiming to leverage complementary modalities such as text and images to infer aspect-level sentiment polarity more accurately.

MABSA typically comprises three subtasks: \emph{Multimodal Aspect Term Extraction} (MATE)~\cite{ref3}, \emph{Multimodal Aspect Sentiment Classification} (MASC)~\cite{ref5}, and \emph{Joint Multimodal Aspect Sentiment Analysis} (JMASA)~\cite{ref7}. Specifically, MATE identifies aspect terms from text--image pairs, MASC predicts sentiment polarity for given aspects, and JMASA performs aspect extraction and sentiment classification jointly, outputting aspect--sentiment pairs.
\begin{figure}
  \begin{center}
  \includegraphics[width=3.5in]{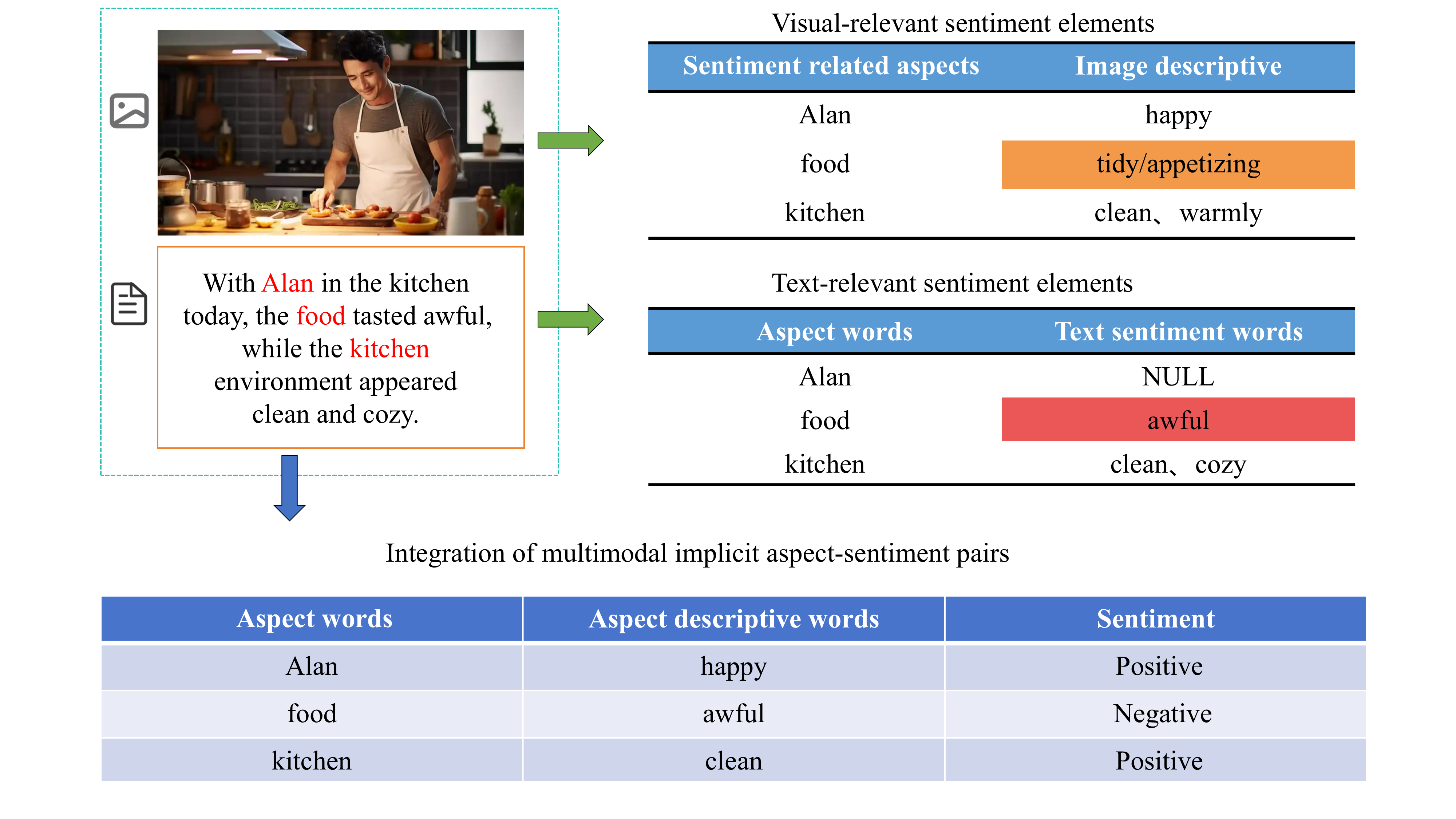}\\
  \caption{An Example of Sentiment Inconsistency in Aspect-Based Sentiment Analysis}
  \label{diagram1}
  \end{center}
\end{figure}

Existing studies primarily improve multimodal fusion by enhancing image--text alignment~\cite{ref10}. Some approaches localize salient visual regions using object detectors~\cite{ref11} and align them with textual tokens, while others focus on global image--text alignment to capture holistic semantics~\cite{ref14}, which may introduce aspect-irrelevant background noise. To mitigate such interference, methods such as TMFN~\cite{ref15} and AoM~\cite{ref17} emphasize aspect-related visual evidence, and Transformer-based models (e.g., CoolNet~\cite{ref18}) further strengthen cross-modal interactions. Complementary strategies exploit similarity constraints to reduce modality noise and semantic bias~\cite{ref19}.

Despite these advances, several challenges remain. 
First, text and images may express inconsistent or even conflicting sentiments toward the same aspect (see Fig.~\ref{diagram1}), yet many methods emphasize cross-modal complementarity while insufficiently modeling consistency--conflict cues, resulting in biased predictions under modality disagreement. 
Second, both modalities contain substantial aspect-irrelevant noise (e.g., redundant text or background regions), and without effective filtering, fused representations become less discriminative. 
Third, modality contributions vary across samples and aspects; thus, rigid fusion strategies may over-rely on a single modality when the other is missing or noisy.

To address these issues, we propose a \emph{Gated Noise-filtered Sentiment-Relevance Interaction} (GNSRI) framework for MABSA. GNSRI employs a gated noise-filtering mechanism to suppress aspect-irrelevant interference and enhance sentiment-related representations, followed by a sentiment-relevance interaction module that jointly models cross-modal consistency and conflict at both fine-grained and global levels. Moreover, an aspect-level adaptive decision fusion strategy dynamically integrates predictions from textual, visual, and cross-modal branches, improving robustness under varying modality dominance. Extensive experiments on two public MABSA benchmarks demonstrate that GNSRI consistently outperforms state-of-the-art methods on MATE, MASC, and JMASA tasks.Our main contributions are as follows:
\begin{itemize}
  \item We propose GNSRI, a novel MABSA framework that suppresses sentiment-irrelevant noise and highlights sentiment-relevant cues via gated filtering and sentiment-relevance interaction.
  \item We introduce an aspect-level adaptive decision fusion strategy to dynamically integrate textual, visual, and cross-modal predictions, enhancing robustness and generalization.
  \item Extensive experiments on two public benchmarks show that GNSRI consistently outperforms state-of-the-art methods across multiple MABSA tasks.
\end{itemize}


\section{Related Work}
Prior work on multimodal aspect-based sentiment analysis (MABSA) largely improves multimodal fusion by enhancing image--text alignment~\cite{ref11}, through multi-granularity contrastive learning (MGIGL)~\cite{ref24} or task-oriented vision--language pretraining (VLP-MABSA)~\cite{ref11}. Despite these advances, existing methods often under-address redundant intra-modal information and semantic/affective inconsistencies between text and image. To this end, we introduce gated noise filtering and sentiment-relevance interaction to suppress modality noise while highlighting both consistent evidence and informative discrepancies for sentiment reasoning.

For aspect-oriented sentiment classification, ESAFN~\cite{ref26} reduces visual noise via entity-sensitive attention, while UnifiedTMSC~\cite{ref29} adopts a prompt-based formulation to simplify aspect handling and improve prediction consistency.

For joint aspect--sentiment extraction (JMASA), representative studies include JML~\cite{ref7}, which jointly optimizes extraction and classification, and AoM~\cite{ref17}, which emphasizes aspect-related semantic/affective cues to reduce irrelevant interference. More recently, EaNet~\cite{ref31} further strengthens extraction through fine-grained aspect--region alignment and cross-modal interference suppression. In contrast, our approach unifies the subtasks within a single framework and explicitly tackles both modality noise and image--text inconsistency via gated noise filtering and sentiment-relevance interaction.

\section{Method Design and Analysis}
\begin{figure*}[t]
    \centering
    \includegraphics[width=\textwidth]{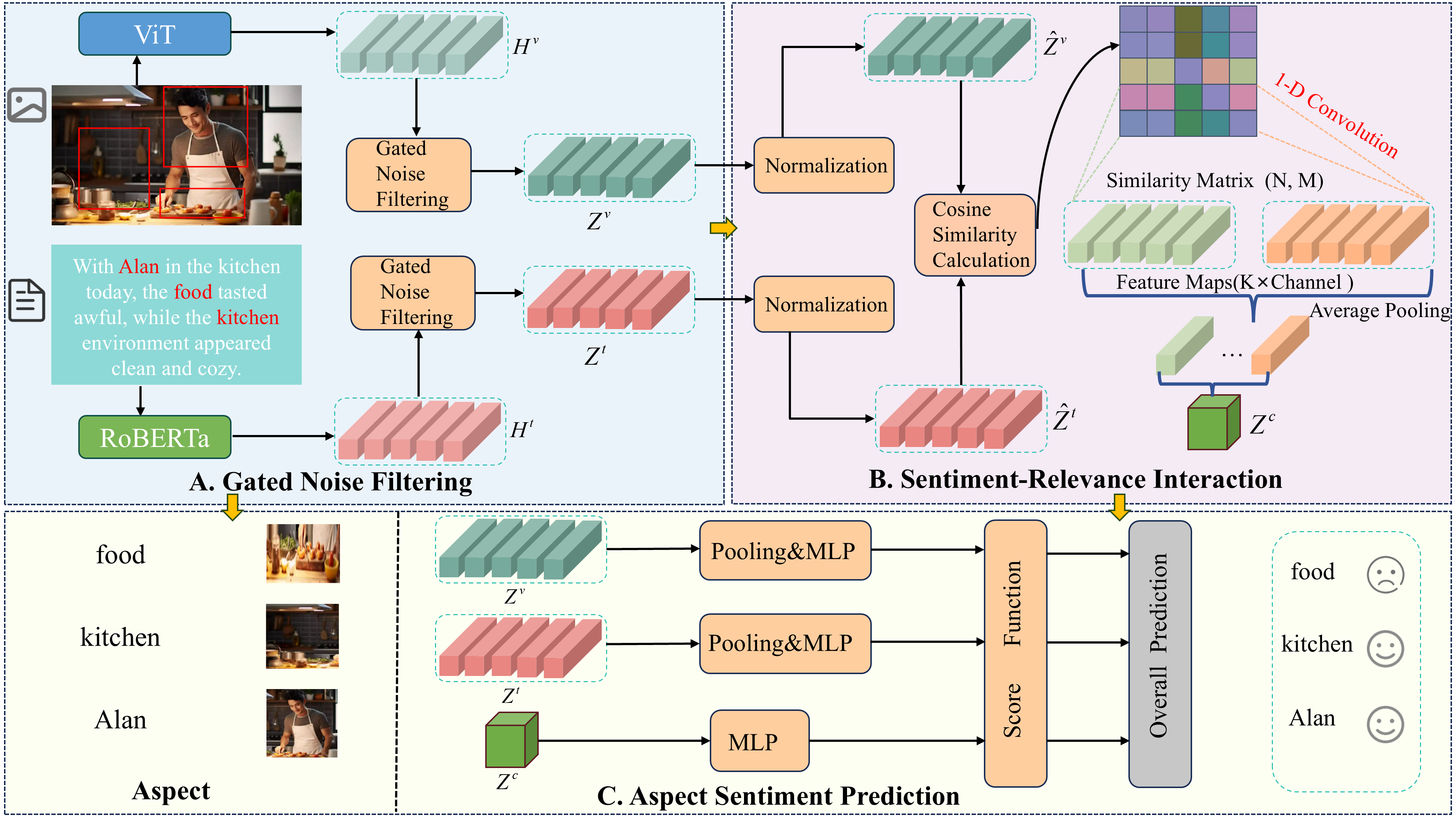}
    \caption{Overall architecture of the proposed GNSRI model. 
    It consists of three main components: (A) Gated Noise Filtering, 
    (B) Sentiment-Relevance Interaction, and 
    (C) Aspect Sentiment Prediction.}
    \label{fig:gnsri_framework}
\end{figure*}

This section provides a detailed description of the GNSRI model, whose overall architecture is illustrated in Fig.~\ref{fig:gnsri_framework}. GNSRI consists of three main components: Gated Noise Filtering, Sentiment-Relevance Interaction, and Aspect Sentiment Prediction. 

\subsection{Task Definition}
Given a multimodal input $x = (x^{T}, x^{V}) \in D$, where $x^{T}$ and $x^{V}$ denote the textual and visual modalities, respectively, the goal is to predict a sequence
\[
Y = \left[(a_{1}, s_{1}), \ldots, (a_{k}, s_{k})\right],
\]
in which each pair consists of an aspect term $a_i$ and its corresponding sentiment polarity $s_i \in \{\text{POS}, \text{NEU}, \text{NEG}\}$, representing positive, neutral, and negative sentiments.

Following prior work, we adopt RoBERTa~\cite{ref32} and ViT~\cite{ref33} to extract contextualized textual and visual representations, denoted as
$H^{t} = \{h^{t}_{1}, \ldots, h^{t}_{n}\}$ and
$H^{v} = \{h^{v}_{1}, \ldots, h^{v}_{n}\}$, respectively.
The visual features are further projected through a multilayer perceptron (MLP) to align with the textual feature space.
Both $H^{t}$ and $H^{v}$ lie in $\mathbb{R}^{n \times d}$, where $d$ is the hidden dimension.

\subsection{Gated Noise Filtering}

The Gated Noise Filtering (GNF) module aims to suppress sentiment-irrelevant noise in both textual and visual modalities while enhancing sentiment-related semantic features. It combines multi-head self-attention with a gating mechanism to refine intra-modal representations.

Specifically, a multi-head self-attention mechanism with $h$ heads is employed to capture contextual features from different subspaces. The output of the $i$-th head is computed as:
\begin{equation}
M_i = \mathrm{Attention}(Q_i, K_i, V_i),
\label{eq:gnf1}
\end{equation}
where $Q_i = XW_i^{Q}$, $K_i = XW_i^{K}$, and $V_i = XW_i^{V}$ are the query, key, and value matrices, respectively. Here, $X \in \mathbb{R}^{n \times d}$ denotes the input sequence, and $W_i^{Q}, W_i^{K}, W_i^{V} \in \mathbb{R}^{d \times d_k}$ are learnable projections with $d_k = d/h$.

While multi-head attention captures contextual dependencies, it may also encode sentiment-irrelevant information. To alleviate this issue, we introduce a gated attention fusion mechanism to filter noisy features. For the $i$-th head, $Q_i$ and $K_i$ are projected into a shared space and fused as:
\begin{equation}
G_i = (Q_i W_{Q}^{G}) \odot (K_i W_{K}^{G}),
\label{eq:gnf2}
\end{equation}
\begin{equation}
U_{Q}^{i} = \sigma(G_i W_{Q}^{M}), \quad U_{K}^{i} = \sigma(G_i W_{K}^{M}),
\label{eq:gnf3}
\end{equation}
where $G_i \in \mathbb{R}^{n \times d_k}$ is the fused representation, $U_Q^i$ and $U_K^i$ are gating masks, and $\sigma(\cdot)$ denotes the activation function.

The gating masks are applied to the query and key representations to suppress sentiment-irrelevant information:
\begin{equation}
\tilde{M}_i = \mathrm{Attention}(Q_i \odot U_{Q}^{i},\, K_i \odot U_{K}^{i},\, V_i),
\label{eq:gnf4}
\end{equation}
where $\tilde{M}_i \in \mathbb{R}^{n \times d_k}$ is the gated attention output of the $i$-th head.

The outputs of all heads are concatenated and combined with a residual connection:
\begin{equation}
f(X) = \mathrm{Concat}(\tilde{M}_1, \ldots, \tilde{M}_h) + X.
\label{eq:gnf5}
\end{equation}

Finally, an MLP with a residual connection is applied to obtain the GNF output:
\begin{equation}
F_{\mathrm{GNF}}(X) = f(X) + \mathrm{MLP}(f(X)).
\label{eq:gnf6}
\end{equation}

The GNF module is applied to both modalities, producing sentiment-relevant representations $Z^{t} = F_{\mathrm{GNF}}(H^{t})$ and $Z^{v} = F_{\mathrm{GNF}}(H^{v})$.

\subsection{Sentiment-Relevance Interaction}

The Sentiment-Relevance Interaction (SRI) module models cross-modal sentiment correlations from both fine-grained and coarse-grained perspectives, encoding consistent and conflicting sentiment cues into a unified relevance representation. At the fine-grained level, it captures interactions between aspect terms and image regions, while at the coarse-grained level, it models sentiment alignment between the entire text and the global image.

To compute cross-modal similarity, textual and visual representations are first normalized to obtain $\hat{Z}^{t}$ and $\hat{Z}^{v}$. The cosine similarity between textual and visual units is then calculated as:
\begin{equation}
s_{ij} = \hat{Z}^{t}_i \cdot \hat{Z}^{v}_j,
\label{eq:sri7}
\end{equation}
where $s_{ij}$ denotes the similarity between the $i$-th textual unit and the $j$-th visual unit.

All similarity scores are organized into a text--image relevance matrix $S = \{s_{ij}\}_{N \times M}$, where $N$ and $M$ denote the numbers of aspect terms and image regions, respectively. The element at $(0,0)$ captures global text--image relevance, while the remaining entries encode aspect--region interactions.

To model local interaction patterns between sequential text and spatial visual features, a one-dimensional convolution is applied along the textual dimension. Given a kernel of height $h$ and width $M$, the convolution is defined as:
\begin{equation}
c_i = \max(0, w \cdot S_{i:i+h-1} + b),
\label{eq:sri8}
\end{equation}
where $w \in \mathbb{R}^{h \times M}$ and $b$ are learnable parameters.

Sliding the kernel with stride 1 produces a feature map:
\begin{equation}
C_i = [c_1, c_2, \ldots, c_{N-h+1}],
\label{eq:sri9}
\end{equation}

To capture relevance patterns at multiple granularities, multiple kernels with different sizes are used, and each feature map is aggregated via global average pooling:
\begin{equation}
\tilde{C}_i = \mathrm{AvgPool}(C_i).
\label{eq:sri10}
\end{equation}

Finally, the pooled features are concatenated to form the cross-modal relevance representation:
\begin{equation}
Z^{c} = \tilde{C}_1 \oplus \tilde{C}_2 \oplus \cdots \oplus \tilde{C}_l,
\label{eq:sri11}
\end{equation}
where $l = k \times o$, with $k$ kernel sizes and $o$ output channels.

\subsection{Aspect Sentiment Prediction}

Given the textual, visual, and cross-modal relevance features, we observe that the contribution of each modality varies across aspects. To integrate modality-specific predictions, we propose a learnable decision fusion mechanism with aspect-level dynamic weighting.

Specifically, modality weights are normalized using Softmax, and the final sentiment prediction for aspect $i$ is computed as:
\begin{equation}
\hat{y}_i = \sum_{m \in \{t,v,c\}} 
\frac{\exp(\omega_{i,m})}{\sum_j \exp(\omega_{i,j})} S_{i,m},
\label{eq:asp12}
\end{equation}
where $m \in \{t, v, c\}$ denotes the text, visual, and cross-modal relevance modalities, respectively. $S_{i,m}$ is the sentiment score predicted by modality $m$, and $\omega_{i,m}$ is a learnable parameter measuring the contribution of modality $m$ to aspect $i$.

\section{Experiments and Analysis}

\textbf{Datasets.}
We evaluate our method on two public multimodal benchmarks, Twitter-2015~\cite{ref28} and Twitter-2017~\cite{ref28}, each consisting of paired text--image samples with aspect terms and aspect-level sentiment labels (positive/negative/neutral). Dataset statistics are summarized in Table~\ref{tab:dataset_statistics}.

\textbf{Evaluation Metrics.}
Following standard settings in prior work~\cite{ref34}, we adopt commonly used metrics to evaluate model performance. For JMASA and MATE, we report Precision (P), Recall (R), and F1 to assess aspect extraction and sentiment prediction. For MASC, we use Precision (P) and F1 for sentiment classification.

\textbf{Implementation Details.}
We use pre-trained RoBERTa and ViT to extract textual and visual features, respectively. Experiments are conducted on Windows with Python~3.10 and PyTorch~1.12.0, running on an NVIDIA RTX~4090 GPU. The learning rate is set to $2\times10^{-5}$, dropout to 0.1, and the hidden dimension to 768.

\begin{table}[htbp]
\centering
\caption{Statistics of Experimental Datasets}
\label{tab:dataset_statistics}
\renewcommand{\arraystretch}{1.2}
\begin{tabular}{ccccccc}
\hline
\multirow{2}{*}{\textbf{Labels}}
& \multicolumn{3}{c}{\textbf{Twitter-2015}}
& \multicolumn{3}{c}{\textbf{Twitter-2017}}\\ \cline{2-7}
& Train & Dev & Test 
& Train & Dev & Test \\
\hline
Negative & 368 & 149 & 113 & 416 & 144 & 168 \\
Neutral  & 1883 & 670 & 607 & 1638 & 517 & 573 \\
Positive & 928 & 303 & 317 & 1508 & 515 & 493 \\
\hline
One aspect     & \multicolumn{3}{c}{2159 (61.65\%)} 
               & \multicolumn{3}{c}{976 (33.54\%)} \\
Mult. aspects  & \multicolumn{3}{c}{1343 (38.35\%)} 
               & \multicolumn{3}{c}{1934 (66.46\%)} \\
Total Aspects  & \multicolumn{3}{c}{3502} 
               & \multicolumn{3}{c}{2910} \\
\hline
\end{tabular}
\end{table}

\subsection{Baseline Models}

To evaluate the effectiveness of GNSRI, we compare it with representative baselines on MATE, MASC, and JMASA, grouped as follows.

\textbf{Text-based methods.}
SPAN~\cite{ref35} is a text-only span-based end-to-end model for joint aspect extraction and sentiment classification. D-GCN~\cite{ref36} is a BERT-based directional GCN that incorporates syntactic dependency information for joint aspect--sentiment prediction. RoBERTa~\cite{ref32} serves as a strong text-only baseline, using Transformer encoding followed by a CRF layer to extract aspect--sentiment pairs.

\textbf{MATE methods.}
RAN~\cite{ref23} aligns textual tokens with visual targets via cross-modal attention for multimodal aspect term extraction. UMT~\cite{ref37} adopts a multimodal Transformer to fuse text and vision for end-to-end multimodal entity recognition. OSCGA~\cite{ref38} converts detected visual objects into BIO-tag sequences and builds fine-grained vision--text alignments to enhance aspect extraction.

\textbf{MASC methods.}
TomBERT~\cite{ref28} injects visual sentiment cues into BERT to align textual representations with visual signals. ESAFN~\cite{ref26} performs cross-modal interaction with attention and gating to suppress visual noise 
and refine sentiment features. CapTrBERT~\cite{ref39} strengthens cross-modal semantics by translating images into auxiliary textual descriptions.

\begin{table*}[t]
\centering
\caption{Performance comparison of different models on the JMASA task. Methods marked with \textsuperscript{*} are from Ref.~\cite{ref16}.}
\label{tab:jmasa_results}
\renewcommand{\arraystretch}{1.1} 
\begin{tabularx}{\textwidth}{l *{6}{>{\centering\arraybackslash}X}} 
\toprule
\multirow{2}{*}{\textbf{Method}} & \multicolumn{3}{c}{\textbf{Twitter-2015}} & \multicolumn{3}{c}{\textbf{Twitter-2017}} \\ 
\cmidrule(lr){2-4} \cmidrule(lr){5-7} 
 & P & R & F1 & P & R & F1 \\ \midrule

SPAN\textsuperscript{*}            & 53.72 & 53.90 & 53.81 & 59.60 & 61.72 & 60.64 \\
D-GCN\textsuperscript{*}           & 58.30 & 58.82 & 58.56 & 64.20 & 64.12 & 64.16 \\
RoBERTa                            & 61.78 & 65.32 & 63.50 & 65.50 & 66.89 & 66.20 \\
UMT-TomBERT\textsuperscript{*}     & 58.42 & 61.30 & 59.80 & 62.28 & 62.42 & 62.35 \\
OSCGA-TomBERT\textsuperscript{*}   & 61.68 & 63.40 & 62.53 & 63.40 & 63.88 & 63.65 \\
OSCGA-collapse\textsuperscript{*}  & 63.10 & 63.68 & 63.39 & 63.52 & 63.50 & 63.51 \\
UMT-collapse\textsuperscript{*}    & 60.95 & 60.40 & 60.67 & 60.80 & 60.04 & 60.42 \\
JML                                & 64.95 & 63.20 & 64.08 & 66.48 & 65.50 & 65.98 \\
VLP-MABSA                          & 65.12 & 68.28 & 66.66 & 66.92 & 69.20 & 68.04 \\
CMMT                               & 64.60 & 68.72 & 66.60 & 67.64 & 69.40 & 68.51 \\
Atlantis                           & 65.60 & 69.24 & 67.37 & 68.62 & 70.30 & 69.45 \\
AoM                                & 67.88 & 69.32 & 68.60 & 68.40 & 71.08 & 69.71 \\
EaNet                              & \underline{70.60} & \underline{71.68} & \underline{71.14} & \underline{71.54} & \underline{72.60} & \underline{72.07} \\ \hline
\textbf{GNSRI (Ours)}              & \textbf{72.54} & \textbf{72.74} & \textbf{72.64} & \textbf{73.60} & \textbf{73.48} & \textbf{73.54} \\
\bottomrule
\end{tabularx}
\end{table*}
\textbf{JMASA methods.}
UMT+TomBERT and OSCGA+TomBERT apply UMT/OSCGA for extraction and TomBERT for classification. UMT-collapse~\cite{ref38} and OSCGA-collapse~\cite{ref40} unify extraction and classification via label-collapsing. JMT~\cite{ref7} explicitly models cross-modal semantic dependencies for joint learning. CMMT~\cite{ref30} introduces multi-level interaction to regulate visual influence on text. VLP-MABSA~\cite{ref11} uses unified vision--language pretraining with task-specific objectives. AoM~\cite{ref17} exploits local sentiment-aware alignment and noise filtering with GCNs. Atlantis~\cite{ref41} extends to text--vision--audio fusion, while EANet~\cite{ref31} adaptively aligns multimodal sentiment features via sentiment-aware attention.

\subsection{Main Results}
\begin{table}[htbp]
\centering
\caption{Performance comparison of different models on the MATE task.Methods marked with \textsuperscript{*} are from Ref.~\cite{ref16}.}
\label{tab:mate_results}
\renewcommand{\arraystretch}{1.2}
\begin{tabular}{lcccccc}
\hline
\multirow{2}{*}{\textbf{Method}}
& \multicolumn{3}{c}{\textbf{Twitter-2015}}
& \multicolumn{3}{c}{\textbf{Twitter-2017}}\\ \cline{2-7}
& P & R & F1 
& P & R & F1 \\
\hline
RAN\textsuperscript{*}  & 80.52 & 81.55 & 81.03 & 90.73 & 90.65 & 90.69 \\
UMT\textsuperscript{*}  & 77.80 & 81.73 & 79.68 & 86.70 & 86.78 & 86.72 \\
OSCGA\textsuperscript{*}& 81.75 & 82.10 & 81.92 & 90.20 & 90.72 & 90.45 \\
JML                     & 83.60 & 81.23 & 82.40 & 92.10 & 90.68 & 91.40 \\
VLP-MABSA               & 83.56 & 87.94 & 85.70 & 90.80 & 92.65 & 91.72 \\
CMMT                    & 83.90 & 88.15 & 86.05 & 92.20 & \underline{93.92} & 93.10 \\
AOM                     & 84.65 & 87.90 & 86.24 & 91.82 & 92.90 & 92.35 \\
Atlantis                & 84.25 & 87.70 & 85.94 & 91.80 & 93.21 & 92.50 \\
EaNet                   & \underline{87.20} & {\bfseries 89.52} & \underline{88.35}
                        & \underline{93.28} & 93.80 & \underline{93.54} \\
\hline
{\bfseries GNSRI(Ours)} & {\bfseries 89.62} & \underline{88.54} & {\bfseries 89.08}
                        & {\bfseries 94.40} & {\bfseries 94.65} & {\bfseries 94.52} \\
\hline
\end{tabular}
\end{table}
This section reports the experimental results of our proposed GNSRI on three tasks, namely MATE, MASC, and JMASA, and compares it against state-of-the-art (SOTA) approaches. The results demonstrate that GNSRI consistently achieves notable performance gains across all tasks, thereby substantiating its effectiveness and superiority.

\textbf{Results on JMASA.} Table~\ref{tab:jmasa_results} summarizes the results on JMASA. \textsc{GNSRI} consistently outperforms all baselines on both datasets; in particular, it improves F1 over EaNet by 1.50\% on Twitter-2015 and 1.47\% on Twitter-2017, demonstrating stronger generalization for jointly extracting aspect terms and their polarities. Multimodal methods also generally surpass text-only models, underscoring the value of visual cues for richer sentiment evidence.

\textbf{Results on MATE.} As shown in Table~\ref{tab:mate_results}, \textsc{GNSRI} demonstrates clear gains on MATE.On Twitter-2015, it improves Accuracy and F1 over EaNet by 2.42\% and 0.73\%, with a minor Recall drop of 0.98\%. On Twitter-2017, \textsc{GNSRI}  surpasses EaNet across all metrics. These results indicate that gated noise filtering and sentiment relevance interaction help better capture aspect-related cues and enhance term recognition in multimodal settings.

\textbf{Results on MASC.}
As reported in Table~\ref{tab:masc_results}, \textsc{GNSRI} achieves the best performance on MASC. On Twitter-2017, it outperforms EaNet by 0.50\% in Accuracy and 0.75\% in F1, suggesting improved generalization for multimodal sentiment classification.

\subsection{Ablation Study}
We ablate GNSRI on JMASA with Twitter-2015/2017 (Table~\ref{tab:ablation}) using four variants: \emph{w/o Img}, \emph{w/o GNF}, \emph{w/o SRI}, and \emph{w/o ADF}. All modules improve performance, and SRI contributes the most: removing it drops Twitter-2015 Acc/F1 from 72.54/72.64 to 69.50/68.96 and Twitter-2017 F1 from 73.54 to 68.78, showing the necessity of modeling cross-modal consistency and conflict. Without GNF, F1 decreases by 2.43/2.40 points on Twitter-2015/2017; replacing ADF with equal-weight fusion further reduces F1 by 1.46/2.18 points. The \emph{w/o Img} variant consistently performs worse, confirming the importance of visual cues. 
Overall, these results demonstrate that GNSRI more robustly handles fine-grained aspect distinctions and weak textual negativity under strong visual dominance by leveraging gated noise filtering, sentiment-related interaction, and adaptive fusion.
\begin{table}[htbp]
\centering
\caption{Performance comparison of different models on the MASC task.Methods marked with \textsuperscript{*} are from Ref.~\cite{ref16}.}
\label{tab:masc_results}
\renewcommand{\arraystretch}{1.2}
\begin{tabular}{lcccc}
\hline
\multirow{2}{*}{\textbf{Method}}
& \multicolumn{2}{c}{\textbf{Twitter-2015}}
& \multicolumn{2}{c}{\textbf{Twitter-2017}} \\ \cline{2-5}
& P & F1
& P & F1 \\
\hline
ESAFN\textsuperscript{*}      & 73.42 & 67.40 & 67.84 & 64.25 \\
TomBERT\textsuperscript{*}   & 77.24 & 71.80 & 70.50 & 68.10 \\
CapTrBERT\textsuperscript{*}  & 77.90 & 73.20 & 70.32 & 70.20 \\
IML       & 78.70 & --    & 72.70 & --    \\
VLP-MABSA & 78.64 & 73.80 & 73.80 & 71.82 \\
CMMT      & 77.90 & --    & 73.84 & --    \\
AoM       & 80.20 & 75.92 & 76.42 & 75.10 \\
Atlantis  & 79.32 & --    & 74.20 & --    \\
EaNet     & \underline{81.60} & \underline{78.42} & \underline{78.10} & \underline{75.50} \\
\hline
{\bfseries GNSRI(Ours)}
          & {\bfseries 82.40}
          & {\bfseries 78.85}
          & {\bfseries 78.60}
          & {\bfseries 76.25} \\
\hline
\end{tabular}
\end{table}
\begin{table}[htbp]
\centering
\caption{Ablation study results of the GNSRI model}
\label{tab:ablation}
\renewcommand{\arraystretch}{1.15}
\begin{tabular}{lcccccc}
\hline
\multirow{2}{*}{\textbf{Method}}
& \multicolumn{3}{c}{\textbf{Twitter-2015}}
& \multicolumn{3}{c}{\textbf{Twitter-2017}} \\ \cline{2-7}
& P & R & F1 
& P & R & F1 \\
\hline
\textbf{GNSRI} 
& \textbf{72.54} & \textbf{72.74} & \textbf{72.64} 
& \textbf{73.60} & \textbf{73.48} & \textbf{73.54} \\
w/o Img 
& 69.70 & 73.20 & 71.41
& 71.72 & 71.64 & 71.68 \\
w/o ADF 
& 70.80 & 71.56 & 71.18 
& 70.90 & 71.82 & 71.36 \\
w/o GNF 
& 70.33 & 70.10 & 70.21 
& 70.18 & 72.12 & 71.14 \\
w/o SRI 
& 69.50 & 68.43 & 68.96 
& 68.21 & 69.35 & 68.78 \\
\hline
\end{tabular}
\end{table}

\subsection{Case Study}
We conduct a qualitative case study on two representative examples from Twitter-2015/2017 (Table~\ref{tab:case_study_single}) to further verify GNSRI. These examples exhibit cross-modal inconsistency and fine-grained aspect-level sentiment shifts, where existing methods are easily biased. In the first example, although the image is overall positive, the text explicitly conveys negative sentiment toward the food and positive sentiment toward the kitchen. AoM misclassifies food as positive, whereas EaNet and GNSRI both correctly distinguish the sentiments of different aspects.In the second example, strong positive visual cues mask a weak negative description of Liam (“a bit hesitant”), while Sophie remains positive. AoM and EaNet again collapse to positive predictions, whereas GNSRI leverages cross-modal evidence to distinguish their sentiments correctly.
\begin{table}[htbp]
\centering
\caption{Case study of different models on the JMASA task}
\label{tab:case_study_single}
\renewcommand{\arraystretch}{1.4}
\setlength{\tabcolsep}{2pt} 
\footnotesize

\begin{tabularx}{\columnwidth}{>{\hsize=0.5\hsize}L >{\hsize=1.25\hsize}L >{\hsize=1.25\hsize}L}
\toprule
\textbf{Image} & 
\includegraphics[width=\linewidth, height=1.6cm, valign=m]{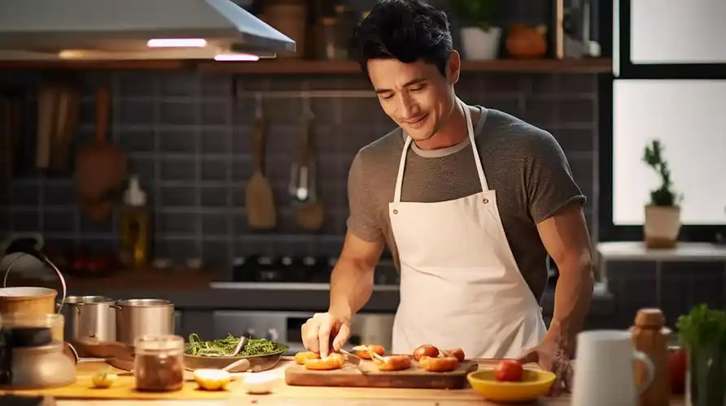} & 
\includegraphics[width=\linewidth, height=1.6cm, valign=m]{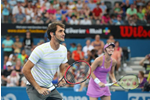} \\ \midrule

\textbf{Text} & 
With Alan in the kitchen today, the food tasted awful, while the kitchen environment appeared clean and cozy. & 
Both \textit{Liam} and \textit{Sophie} looked excited during the match, though Liam's footwork seemed a bit hesitant at times. \\ \midrule

\textbf{GT} & 
(Alan, POS)\par (food, NEG)\par (kitchen, POS) & 
(Liam, NEG)\par (Sophie, POS) \\ \midrule

\textbf{AoM} & 
(Alan, POS) \textcolor{purple}{\ding{52}}\par (food, POS) \textcolor{red}{\ding{56}}\par (kitchen, POS) \textcolor{purple}{\ding{52}} & 
(Liam, POS) \textcolor{red}{\ding{56}}\par (Sophie, POS) \textcolor{purple}{\ding{52}} \\ \midrule

\textbf{EaNet} & 
(Alan, POS) \textcolor{purple}{\ding{52}}\par (food, NEG) \textcolor{purple}{\ding{52}}\par (kitchen, POS) \textcolor{purple}{\ding{52}} & 
(Liam, NEU) \textcolor{red}{\ding{56}}\par (Sophie, POS) \textcolor{purple}{\ding{52}} \\ \midrule

\textbf{GNSRI} & 
(Alan, POS) \textcolor{purple}{\ding{52}}\par (food, NEG) \textcolor{purple}{\ding{52}}\par (kitchen, POS) \textcolor{purple}{\ding{52}} & 
(Liam, NEG) \textcolor{purple}{\ding{52}}\par (Sophie, POS) \textcolor{purple}{\ding{52}} \\ 
\bottomrule
\end{tabularx}
\end{table}

\section{Conclusion}
This paper proposes GNSRI, a multimodal aspect-level sentiment analysis framework built upon gated noise filtering and sentiment-related interaction. 
First, a gated noise filtering module is introduced to suppress irrelevant disturbances in both text and images, thereby enhancing sentiment-indicative semantic cues. 
Second, a sentiment-related interaction module explicitly models cross-modal consistency and conflict signals to improve the usability and discriminability of multimodal sentiment evidence. 
Third, a learnable adaptive decision fusion mechanism dynamically re-weights the text, image, and cross-modal branches at the aspect level, leading to more stable and robust predictions. 
Experimental results on the public Twitter-2015 and Twitter-2017 benchmarks demonstrate that GNSRI consistently outperforms existing methods across the MATE, MASC, and JMASA settings, validating its effectiveness for multimodal aspect-level sentiment analysis.

\bibliographystyle{IEEEtran}
\bibliography{icme2025references}

\end{document}